\documentclass[final, 3p, numbers, times]{elsarticle}

\usepackage{amssymb}
\usepackage{multirow}
\usepackage{setspace}
\usepackage{amsthm}
\usepackage{amsmath}
\usepackage[centerlast]{subfigure}
\usepackage{float}
\usepackage{stfloats} 
\usepackage{algorithm}
\usepackage{algorithmic}

\journal{Reliability Engineering and System Safety}
\DeclareMathOperator{\C}{C}
\DeclareMathOperator{\R}{R}
\DeclareMathOperator{\E}{E}
\DeclareMathOperator{\F}{F}
\DeclareMathOperator{\Y}{Y}

\DeclareMathOperator{\T}{T}
\DeclareMathOperator{\CO}{CO}
\begin{document}

\begin{frontmatter}



\title{Shape invariant model approach for functional data analysis in uncertainty and sensitivity studies}


\author[UT,IFP]{Ekaterina Sergienko}
\author[UT]{Fabrice Gamboa}
\author[IFP]{Daniel Busby}

\address[UT]{Universit\'e Paul Sabatier, IMT-EPS, 118, Route de Narbonne, 31062, Toulouse, France}
\address[IFP]{IFP Energies Nouvelles, 1-4 avenue de Bois-Pr\'eau, 92582, Rueil-Malmaison, France}

\begin{abstract}
Dynamic simulators model systems evolving over time. Often, it operates iteratively over fixed number of time-steps. The output of such simulator can be considered as time series or discrete functional outputs. Metamodeling is an effective method to approximate demanding computer codes. Numerous metamodeling techniques are developed for simulators with a single output. Standard approach to model a dynamic simulator uses the same method also for multi-time series outputs: the metamodel is evaluated independently at every time step. This can be computationally demanding in case of large number of time steps. In some cases, simulator outputs for different combinations of input parameters have quite similar behaviour. In this paper, we propose an application of shape invariant model approach to model dynamic simulators. This model assumes a common pattern shape curve and curve-specific differences in amplitude and timing are modelled with linear transformations. We provide an efficient algorithm of transformation parameters estimation and subsequent prediction algorithm. The method was tested with a $\CO_{2}$ storage reservoir case using an industrial commercial simulator and compared with a standard single step approach. The method provides satisfactory predictivity and it does not depend on the number of involved time steps.
\end{abstract}


\end{frontmatter}

% \linenumbers

\section{Introduction}
\label{}
Simulation models are used nowadays in many industrial applications to predict and analyze the behaviour of complex systems. A simulator is a complex computer code embedding the physical laws governing the physical system under investigation. The input of such simulators can be adjustable or uncontrollable parameters which are only partially known and thus are affected by uncertainty. Uncertainty analysis of numerical experiments is used to assess the confidence of the model outcomes which are then used to make decisions \cite{Roc2008}. Here, we focus on a particular type of dynamic simulators that are used to make predictions in the future. These simulators are based typically on finite element/finite difference codes used for instance for the simulation of flows and transfers in porous media. Industrial applications using this type of simulators are for instance hydrocarbons reservoir forecasting and carbon dioxide ($\CO_{2}$) underground storage \citep{segses11, ecmor10}. Such applications involve very complex numerical codes with a large number of inputs and with high uncertainty derived from an incomplete knowledge of subsurface formation \citep{subbey2004}.  The uncertainty on the simulator output is usually assumed to be mostly due to the propagation of the uncertainty on the input. Nevertheless, modelling errors can also play a role. 
\par 
The simulator models a multi-phase 3-D fluid flow in heterogeneous porous media, operating over fixed number of time-steps. The typical output of such simulators consists of a sequence of outputs at different time-steps. Therefore, it represents time series related, for instance, to a recovery rate for a given well or a group of wells. It can be also a spatial output such as a pressure map or a saturation map also evolving with time. Here we focus on 1D time series output which can be typically measured in a well. 
\par
 Let us consider the output of a simulator as a deterministic function $\Y(t)=\F(\bar{x},t)$, where $\bar{x} \in \Omega \subset \mathbb{R}^{d}$ is a configuration of preselected input parameters and $0<t<T$ refers to a time step. $\Y(t)$ is a time dependent output, e.g. oil production rate or reservoir pressure. 
\par
The function $F:\mathbb{R}^d\times[0, T]\rightarrow \mathbb{R}$ models the relationship between the input and the output of the numerical simulator. Methods such as Monte Carlo ones can be used to propagate uncertainty. However, each model evaluation being generally very time-consuming this approach can be unpractical in industrial applications. Moreover, the probability distribution of the input may vary as new data comes in or within a process of dynamic data assimilation. Therefore, a response surface approach (also referred to a metamodel approach) can be more useful in such applications \citep{ecmor10, busby2008, articlebusby07}.
\par
The advantage of a metamodel is that is fast to evaluate and it is designed to approximate the complex computer code based on a limited number of simulator evaluations. These evaluations of the simulator are taken at some well chosen input configurations also called the training set or the experimental design. Numerous experimental designs have been proposed and many of them are quite sophisticated. Here, we use Latin Hypercube designs \cite{Mckay79} for their good space filling properties. Usually, it can be coupled with other criteria such as the maximin design (maximum minimum distance) \citep{Sacks1989, Santner2003}. In this work, we focus on Gaussian process (GP) based metamodels also known as \textit{kriging} \cite{Matheron1963, Sacks1989, Santner2003}. 
\par
The aim of this work is to propose a new method to address multi-time series outputs. For such dynamic simulator a standard approach assumes a single step GP model for each time step. This basic approach can be computationally intensive for a high number of time steps and when the construction of a single GP model is time consuming (i.e. when the size of the training set and the number of variables is large). Therefore, the procedure can become unpractical in some industrial applications. 
\par
The problem of metamodeling for dynamic simulators was recently investigated by different authors and several principal approaches can be distinguished. A first possible approach is GP metamodelling considering the time steps as the model additional input parameter \cite{ecmor10, conti2009, Qian2007}. This approach is easy to implement, however if we want to take into account all the information from an experimental design at every time steps, the size of new experimental design is multiplied by the number of time steps.  It results to matrices of high dimensions  that can lead to some numerical problems in case of large size of original experimental design or in case of high density of the steps in the time scale. Conti et al. (2009)  \cite{Ohagan2009} developed an iterative approach to build a model of dynamic computer code, assuming that the model output at a given time step depends only on the output at the previous time step. To reduce the dimensionality of the problem, we can also represent the given observations as truncated functions by some selected basis functions with the following GP modelling for the coefficient of the selected representation. Bayarri et al. (2007) \cite{bayarri2007} introduced wavelet decomposition. Campbell et al. (2006) \cite{campbell2006}, Higdon et al. (2008) \cite{higdon2008} and Lamboni et al. (2009) \cite{lamboni2009} suggested application of principal component decomposition. Auder et al. (2010) \cite{auder2010} extended this approach by preliminary classification. 
 \par
In this work, we propose a new functional approach involving a combination of Shape Invariant Model (SIM) approach and Gaussian Process modelling.  
Considering $J$ time steps and a training set $\mathbf{X}^{n}=\left\{\bar{x}_{1},..,\bar{x}_{n}\right\}$ 
the simulator output consist then in a set of curves:
\begin{center}
$\mathbf{Y}^{n}=\left\{Y_{i,j}=F(\bar{x}_{i}, t_{j}), 1\leq i \leq n, 1\leq j \leq J \right\}$ . 
\end{center}
\par
In our practical example (the $\CO_{2}$ storage case study) we have observed that usually these curves have quite similar behaviour. So that, we assume that there is a mean pattern from which all the curves can be deduced by a proper parametrical transformation. The shape invariant model assumes that the curves have a common shape, which is modelled nonparametrically. Then curve-specific differences in amplitude and timing are modelled with the linear transformations such as shift and scale \citep{lawton1972, brumback2002, izem2007}. Hence, we consider the following model:
\begin{equation}
\label{SIM1}
Y_{i,j} = \alpha^{*}_{i} \ \ f(t_{j}-\theta^{*}_{i})+v^{*}_{i}+\sigma_{i}^{*}  \varepsilon_{ij} ,\ \ 1 \leq i \leq n , 1\leq j \leq J
\end{equation}
where $t_j$ stands for observation times and $\boldsymbol{\theta}^{*}=\left\{\theta_{i}, 1\leq i \leq n\right\}$, $\boldsymbol{v}^{*}=\left\{v_{i}, 1\leq i \leq n\right\}$, $\boldsymbol{\alpha}^{*}=\left\{\alpha_{i}, 1\leq i \leq n\right\}$ $\in \mathbb{R}^{n}$ are vectors of transformation parameters corresponding to horizontal and vertical shifts and scales of the unknown function $f$. 
\par
Without loss of generality we assume that the simulator provides the data at some constant time intervals. 
Time $t$ is then equispaced in $\left[0,\T \right]$ with $J$ time steps. The unknown errors $\sigma_{i}^{*} \varepsilon_{i,j} $ are independent zero-mean random variables with variance ${\sigma_{i}^{*}}^{2}$. Here, we can also assume that ${\sigma_{i}^{*}}^{2}=1$ without loss of generality.

\par The approach proposed in this work for the functional outputs modeling combines an efficient estimation of the transformation parameters $(\boldsymbol{\alpha^{*}, \theta^{*}, v^{*} })$ and a subsequent GP modeling for these parameters that can be used to predict new curves without running the simulator. %The method is then applied to uncertainty analysis and risk analysis of $\CO_{2}$ storage problems involving predictions over several hundreds years of simulation.

\par The paper is organized as follows. Section 2 presents the basics of GP modeling and the model validation criteria. Section 3 describes the method of efficient estimation  of transformation parameters in the shape invariant model. The method is illustrated with an example on an analytical
 function. In Section 4 we present the forecast algorithm for dynamic simulators. Section 5 provides the practical application of the algorithm with a $\CO_{2}$ storage reservoir case.

\section{\label{sec:GP}GP based metamodeling}
The method proposed in this paper is based on a combination of a shape invariant model and a Gaussian process metamodel.  In this section, we recall the basics of GP modeling or \textsl{kriging}. 
\par
The idea of modeling an unknown function by a stochastic process was introduced in the field of geostatistics by Krige in the 1950's \cite{krig} and formalized in 1960's by Matheron (1963) \cite{Matheron1963}. Later  Sacks et al. (1989) \cite{Sacks1989} proposed the use of kriging for prediction and design of experiments. The theory and the algorithms are formalized in \cite{Sacks1989}, \cite{Welch1992} and \cite{Santner2003}.
\par
Consider the output of a simulator as an unknown deterministic function $\F(\bar{x}) \in \mathbb{R}$, where $\bar{x} \in \Omega \subset \mathbb{R}^{d}$ is a specified set of selected input parameters. The function $F$ is only known in predetermined design points: $\mathbf{X}^{n}=\left\{\left({\bar{x}_{k}},{\F_{k}=\F(\bar{x}_{k})}\right), 1\leq k \leq n \right\}$. 
The objective is to predict the function $\F_{0} = \F(\bar{x}_{0})$ for some new arbitrary input  $\bar{x}_{0}$. The function is modeled as a sample path of a stochastic process of the form:
\begin{equation}
\label{GP1}
\widetilde{\F}(\bar{x})=\sum^{m}_{j=1}h_{j}(\bar{x})\cdot \beta_{j}+Z(\bar{x})={\boldsymbol{\beta}}^{\top}\mathbf{h}(\bar{x})+Z(\bar{x})
\end{equation} 
where:
\begin{itemize}
	\item  $\boldsymbol\beta^{\top}\mathbf{h}(\bar{x})$ is the mean of the process and corresponds to a linear regression model with preselected given real-valued functions $\mathbf{h}=\left\{h_{i}, 1 \leq i \leq m \right\}$. Here, we only consider the case $\mathbf{h}=\mathbf{1}$.
	\item  $Z(\bar{x})$ is a centered Gaussian stationary random process. It is defined by its covariance function:
	$\C(\bar{x},\bar{y})=\E\left[Z(\bar{x})Z(\bar{y})\right]=\sigma^{2} \R(\bar{x},\bar{y})$. $\R(\bar{x}, \bar{y})$ is the correlation function and $\sigma^{2}=\E[Z(\bar{x})^{2}]$ denotes the process variance. Stationarity condition assumes: $\R(\bar{x},\bar{y})=\R(\left|\bar{x}-\bar{y}\right|)$, where $\left|\bar{x}-\bar{y}\right|$ denotes the distance between $\bar{x}\in \Omega$ and $\bar{y}\in \Omega$.
\end{itemize}
Numerous families of correlation functions have been proposed in the literature. We use here Gaussian correlation function, the special case of the power exponential family. The power exponential correlation function is of the following form:
\begin{equation}
\label{expcorr}
\R(\bar{x},\bar{y})=\exp\left(-\sum^{d}_{j=1}\frac{\left(x_{j}-y_{j}\right)^{p_{j}}}{\theta_{j}}\right)=\exp\left(-\sum^{d}_{j=1}\frac{d_{j}^{p_{j}}}{\theta_{j}}\right)
\end{equation}
where  $d_j=|x_{j}-y_{j}|$, $0<p_{j}\leq 2$ and $\theta_{j}>0$. The hyperparameters $\left( \theta_{1},.., \theta_{d} \right)$ stands for correlation lengths which affect how far a sample point's influence extends. A high $\theta_{i}$ means that all points will have a high correlation ($\F(x_{i})$ being similar across our sample), while a low $\theta_{i}$ means that there are significant difference between the $\F(x_{i})$'s \citep{forr2009}. The parameters $p_{j}$ corresponds to the smoothness parameters. These parameters determine mean-square differentiablity of the random process $Z(x)$. For $p_{j}=2$ the process is infinitely mean-square differentiable and the correlation function is called Gaussian correlation function. Hence, Gaussian correlation function is infinitely mean-square differentiable and it leads to a stationary and anisotropic process $Z(x)$ \citep{Santner2003, Welch1992}.
Regardless the choice of a correlation function, the estimation of hyperparameters $\left( \theta_{1},.., \theta_{d} \right)$  is crucial for reliable prediction. We are using maximum likelihood estimation algorithm \cite{Santner2003} that we will discuss later.

\par The experimental design points are selected in order to retrieve most information on the function at the lowest computational cost. The number of design points for a reliable response surface model depends on the number of inputs and on the complexity of the response to analyze \citep{Mckay79, Santner2003}.  Latin Hypercube Designs (LHD) provides a uniform coverage of the input domain. If we wish to generate a sample of size $\mathbf{n}$, first, we partition the domain of each variable in $\mathbf{n}$ intervals of equal probability. Then, we randomly sample $\bar{x}_{1}$ according to the distribution of each of the $\mathbf{n}$ intervals. Further, for each of the $\mathbf{n}$ values for $\bar{x}_{1}$, we randomly select one interval to sample for $\bar{x}_{2}$ , so that only one sample of $\bar{x}_{2}$ is taken in each interval. We continue the process of a random sampling without replacement until all the variables have been sampled. As a result we generate a sample where each of $d$ inputs is sampled only once in each of $\mathbf{N}$ intervals. Latin hypercube designs have been applied in many computer experiments since they were proposed by Mckay et al., (1979) \citep{Mckay79}.

\par In this work, we use modified version of LHD - maximin LHD. It is based on maximizing a measure of closeness of the points in a design $\mathbf{D}^{n}$:
\begin{equation*}
\max_{\text{design}\ \ \mathbf{D}^{n}} \min_{\bar{x}_{1},\bar{x}_{2} \in \mathbf{D}^{n}} d(\bar{x}_{1},\bar{x}_{2})
\end{equation*}
It can guarantee that any two points in the design are not "too close". Hence, the design points are uniformly  spread over the input domain.

\par Consequently, when we have the experimental design  $\mathbf{X}^{n}=\left( \bar{x}_{1},..,\bar{x}_n\right )$ and the observation data  $\mathbf{Y}^{n}=\left(F(\bar{x}_{1}),..,F(\bar{x}_{n})\right)$ the multivariate  distribution according to the model \eqref{GP1} for the Gaussian correlation function can be expressed as:
\begin{equation*}
\left( \begin{array}{c} Y_{0}\\ \mathbf{Y}^{n} \end{array} \right) \sim \mathcal{N}_{1+n} \left[ \left( \begin{array}{c}
\mathbf{h}^{\top}(\bar{x}_{0}) \\ \mathbf{H} \end{array} \right) \boldsymbol{\beta},\  \ \boldsymbol{\sigma}^{2} \left( \begin{array}{cc}
1 & {r}^{\top}(\bar{x}_{0}) \\
{r}(\bar{x}_{0})& {\mathbf{R}} \end{array} \right)\right] ,
\end{equation*}
where
${\mathbf{R}}=(\R(\bar{x}_{i},\bar{x}_{j}))_{1\leq i,j\leq n}\in \mathbb{R}^{n\times n}$ is the correlation matrix among the observations; ${\mathbf{r}}(\bar{x}_{0})=\left(\R(\bar{x}_{1},\bar{x}_{0}),..,\R(\bar{x}_{n},\bar{x}_{0})\right)^{\top}\in \mathbb{R}^{n}$ is the correlation vector between the observations and the prediction point;
$\mathbf{h}^{\top}(\bar{x}_{0})=\left(h_{j}(\bar{x}_{0})\right)_{1\leq j \leq m}\in \mathbb{R}^{m}$ is the vector of regression function at the prediction point $\bar{x}_{0}$ and
 $\mathbf{H}=\left(h_{j}(\bar{x}_{i})\right)_{1\leq i \leq n, 1 \leq j \leq m}\in \mathbb{R}^{n\times m}$ is the matrix of regression functions at the experimental design. The parameters $\boldsymbol\beta$ and $\boldsymbol\sigma$ are unknown.

\par Considering the unbiasedness constraint, the parameter $\boldsymbol\beta$ is replaced by the generalized least squares estimate  $\widehat{\boldsymbol{\beta}}$ in \eqref{GP1}. Here, $\widehat{\boldsymbol{\beta}}$ is of the following form:
$\widehat{\boldsymbol\beta}=\left(\mathbf{H}^{\top} \hat{\mathbf{R}}^{-1}\mathbf{H}\right)^{-1}\mathbf{H}^{\top} \hat{\mathbf{R}}^{-1}\mathbf{Y}^{n}$.
Assuming that the correlation function is the Gaussian correlation function, the prediction is therefore given by:
\begin{equation}
\widehat{F}(\bar{x}_{0})=\mathbf{h}^{\top}(\bar{x}_{0}) \cdot \hat{\boldsymbol\beta}+\hat{\textbf{r}}(\bar{x}_{0}) \ \ \hat{\mathbf{R}}^{-1}\left(\mathbf{F}^{n}-\mathbf{H} \cdot \hat{\boldsymbol\beta}\right)
\end{equation}
\par
The hyperparameters $ \boldsymbol{\theta} = (\theta_{1},..,\theta_{m})$ and the process variance $\boldsymbol{\sigma}^{2}$  are estimated by Maximum Likelihood (MLE). Using the multivariate normal assumption, the MLE for $\boldsymbol{\sigma}^{2}$ is:
\begin{equation}
\hat{\boldsymbol{\sigma}}^{2}=\frac{1}{m}\left(\mathbf{Y}^{n}-\mathbf{H}\  \hat{\boldsymbol{\beta}}\right) \  \mathbf{R}^{-1}\left(\mathbf{Y}^{n}-\mathbf{H}\  \hat{\boldsymbol{\beta}}\right) 
\end{equation}
Knowing the estimations for $\hat{\boldsymbol{\beta}}$ and $\hat{\boldsymbol{\sigma}}^{2}$, the coefficients $\boldsymbol\theta$ are estimated by maximizing the log likelihood:
\begin{equation}
\label{mle}
l\left(\hat{\boldsymbol{\beta}}, \hat{\boldsymbol{\sigma}}^{2}, \boldsymbol{\theta} \right) =-\frac{1}{2}\left[m \log \hat{\boldsymbol{\sigma}}^{2}(\boldsymbol\theta)+\log(det(\mathbf{R}(\boldsymbol\theta))+m \right]
\end{equation}
The function \eqref{mle} depends only on $\boldsymbol\theta$.
\par After having estimated all the model parameters, we now need to validate the model. In this work for estimation of prediction accuracy of the model, we use the predictivity index, $Q_2$, and root mean squared error, $RMSE$ . The predictivity index is calculated basing on cross validation and it has the following form:
\begin{equation}
RMSE:={\sqrt{\sum^{n}_{i=1}\frac{(\hat{S}_{X/x_{i}}-F(x_{i}))^{2}}{n}}} \ \ \  \
Q_{2}:=1-\frac{\sum_{i=1}^{n}\left(\hat{S}_{X/x_{i}}-F(x_{i})\right)^{2}}{\sum_{i=1}^{n}\left(F(x_{i})-\tilde{F}\right)^{2}}
\end{equation}
$\hat{S}_{X/x_{i}}$ is the kriging model computed using all  the design points $\mathbf{X}^{n}$ excepting $\bar{x}_{i}$ and $\tilde{F}$ is the mean of $F(\bar{x}_{i})$. 

\par The closer $Q_{2}$ is to $1$ or $RMSE$ is to $0$, the higher is the model predictivity. 
These criteria can be also  calculated on a separate validation test data by performing additional simulations. It provides higher accuracy measure though it requires additional time costs.

\section{Shape Invariant Model}
\label{SIMsec} 
\par In this section, we discuss the shape invariant model representation and the procedure for efficient parameters estimation.
\par
The shape invariant model was introduced by Lawton et al. (1972) \cite{lawton1972}. The model assumes that we are working with a set of curves that have a common shape function that is modeled nonparametrically. The deformation of this function is modeled parametrically by choosing a proper parametrical transformation. We consider a class of linear transformations only. These parameters can be normally interpreted as shift in timing or scale in amplitude. For this reason, shape invariant model is widely applied to study periodic data such as temperature annual cycle \citep{vimond2007} or traffic data analysis \citep{Fgamb2007}. Indeed, in these cases there is always some variability in time cycles or amplitude. The model has also been used to study biological data , where the departure from the pattern can be caused by a different environmental conditions \citep{izem2007, brumback2002}.  In this work, we use the model to propagate uncertainty on a reservoir simulator output data. For example, we consider a reservoir pressure during $\CO_{2}$ injection, then shift in time is caused by different moment of stopping injection. So that, by applying shape invariant model, we can study the influence of the model input parameters on the overall shapes of the selected output.
\par
\begin{figure}[H]
\centering
\subfigure[Horizontal Shift]{
	\includegraphics[height=0.3\textwidth]{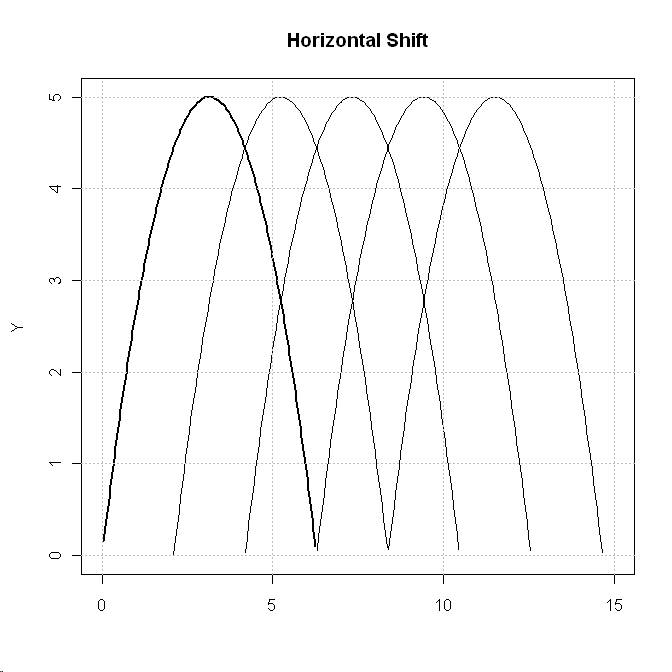}
	\label{fig:horiz}}
\subfigure[Vertical Shift]{
	\includegraphics[height=0.3\textwidth]{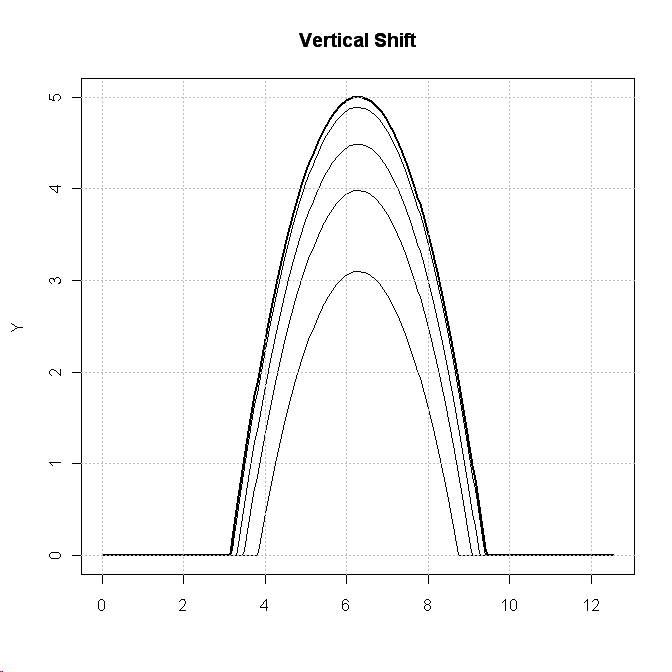}
	\label{fig:verticsh}}
\subfigure[Vertical Scale]{
	\includegraphics[height=0.3\textwidth]{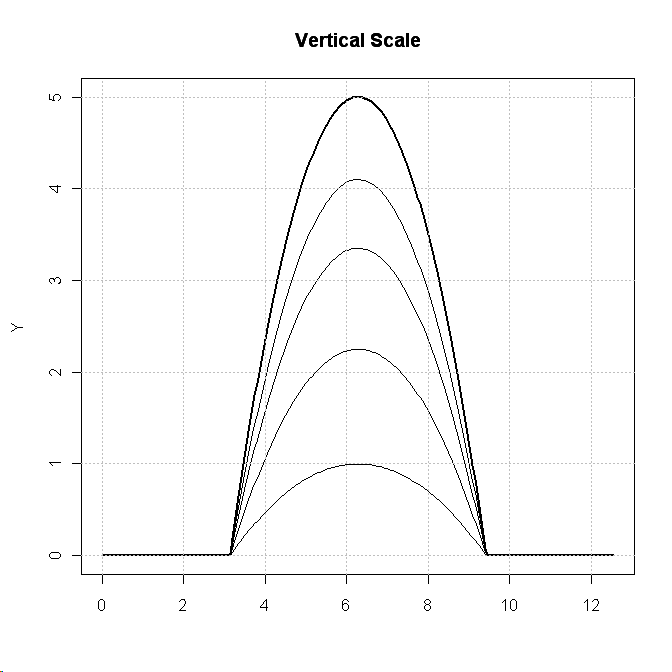}
	\label{fig:verticsc}}
\caption{\label{fig:transfex}Parametrical transformation examples}
\end{figure}

 In figure (\ref{fig:transfex}) we display three possible transformations that we consider in our work:
horizontal shift \ref{fig:horiz}, vertical shift \ref{fig:verticsh} and vertical scaling \ref{fig:verticsc}. The bold line represent original pattern shape. 

\par We are interested in the common shape as well as in the efficient estimation of transformation parameters. So that, we can reproduce any curve and align it to the pattern. Moreover, we can make a prediction of a possible new curve for an input configuration $\bar{x}_{0}$ by modeling the transformation parameters.

\par As already mentioned, for an experimental design $\mathbf{X}^{n}=\left\{ \bar{x}_{1},..,\bar{x}_{n} \right\}$ we have a set of observations: \par $\mathbf{Y}^{n}= \left\{Y_{i,j}=\F(\bar{x}_{i},t_{j}),1\leq i\leq n,\ \ 1\leq j\leq J  \right\}$, where $\bar{x}_{i}, \ \ 1\leq i\leq n$ is the set of preselected input parameters and $t_{j}, \ \ 1\leq j\leq J $ refers to the time sample. So that, {$Y_{i, j}$} is the $j^{th}$ observation on the $i^{th}$ experimental design unit, with $1 \leq i \leq n$ and $1 \leq j \leq J$. Thereby, the idea is to find a general pattern curve that makes possible subsequent transformation of any curve to this selected pattern with properly adjusted transformation parameters.  
\par We focus here on linear transformations. Thus, the model structure may be written as:
\begin{equation}
Y_{i,j} = \alpha^{*}_{i}\cdot f(t_{j}-\theta^{*}_{i})+v_{i}^{*}+{ \sigma^{*} }^{2} \cdot \varepsilon_{ij}
\label{eq:sim}
\end{equation}

where $t_{j}, \ \ 1\leq j\leq J $ are observation times which are assumed to be known and equispaced in the interval $[0, T]$. The vector of parameters: $(\boldsymbol{\alpha^{*}, \theta^{*} , v^{*}})=(\alpha_{1},..,\alpha_{n},\theta_{1},..,\theta_{n},v_{1},..,v_{n})$ is unknown as well as the pattern function $f$. The errors $\varepsilon_{ij}$ are i.i.d. with a normal distribution, $(i,j)\in \{1,..,n\}\times \{1,..,J\}$. It characterizes observation noise. Without loss of generality we may assume that ${\sigma^{*}}^2=1$.  The variance does not affect the parameters estimation procedure and the method still works for a variance ${\sigma^{*}}^2$. The function $f$ is assumed to be $2\pi$-periodic \citep{vimond2007, Fgamb2007}.
\par
In this section we will provide the algorithm for efficient estimation of transformation parameters  under unknown function pattern $f$. Since the functional pattern $f$ is unknown, the pattern is replaced by its estimate. So that, it seems natural to study the problem (\ref{eq:sim}) in a semi-parametric framework: the transformation shifts and scales are the parameters to be estimated, while the pattern stands for an unknown nuisance   functional parameter. We use an M-estimator built on the Fourier series of the data. Under identifiability assumptions it is possible to provide a consistent algorithm to estimate $(\boldsymbol{ \alpha^{*}, \theta^{*}, v^{*}})$ when $f$ is unknown. The algorithm is described in details in the following subsection with an illustration on an analytical function example.

\subsection{Model Assumptions}
Consider $\mathbf{Y}^{n}=F(\mathbf{X}^{n},\boldsymbol{t})=\{F(\bar{x}_{i}, t_{j}), \ \ 1 \leq i \leq n, 1 \leq j \leq J  \}$ is $(n \times J)$ matrix of observations. We model these observations in the following way: 
\begin{equation}
\label{eq:Y}
Y_{i,j} = \alpha^{*}_{i}\cdot f(t_{j}-\theta^{*}_{i})+v_{i}^{*}+ \varepsilon_{ij}, \ \ 1 \leq i \leq n, 1 \leq j \leq J
\end{equation}

where $f:\mathbb{R}\rightarrow \mathbb{R}$ is an unknown $2\pi$-periodic continuous function, $\boldsymbol{\theta}^{*}=(\theta^{*}_{1},...,\theta^{*}_{n})$, $\boldsymbol{\alpha}^{*}=(\alpha^{*}_{1},...,\alpha^{*}_{n})$, $\boldsymbol{v}^{*}=(v_{1},...,v_{n})$  $\in \mathbb{R}^{n}$ are unknown parameters, $\varepsilon_{ij}$ is a Gaussian white noise with variance equal to 1. The time period is translated in such a way that: $[0, \T[ \rightarrow [0, 2\pi[$, therefore $t_{i}=\frac{j-1}{J}2\pi, \ \ j=1,..,J $ are equispaced in $[0, 2 \pi[$.
\par
The objective is to estimate the horizontal shift $\boldsymbol{\theta}^{*}=(\theta^{*}_{1},...,\theta^{*}_{n})$, the vertical shift  $\boldsymbol{v}^{*}=(v_{1},...,v_{n})$ and the scale parameter $\boldsymbol{\alpha}^{*}=(\alpha^{*}_{1},...,\alpha^{*}_{n})$ without knowing of the pattern $f$.  Fourier analysis is well suited for the selected structure of the model. Indeed, this transformation is  linear and  shift invariant. Therefore, by applying a discrete Fourier transform to \eqref{eq:Y} and supposing $J$ is odd, the model becomes:
\begin{equation}
\label{eq:fft}
d_{kl}= \left\{ 
\begin{array}{ll}
\alpha^{*}_{k} e^{-il\theta_{k}^{*}} c_{l}(f) +w_{kl}, \ \  {1 \leq k \leq n , \ \ 0 <\left|l\right| \leq (J-1)/2 } \\
\alpha^{*}_{k}  c_{0}(f) +v^{*}_{k}+w_{k0},\  \   {1 \leq k \leq n , \ \ l=0}
\end{array} \right.
\end{equation}
 where $c_{l}(f)=\frac{1}{J}\sum_{m=1}^{J} f(t_{m}) e^{-2i\pi \frac{ml}{J}},\ \ ( \scriptstyle { \left|l\right| \leq (J-1)/2} \textstyle )$ 
 are the discrete Fourier coefficients and $w_{kl},\ \ ( \scriptstyle {1 \leq k \leq n , \ \ \left|l\right| \leq (J-1)/2}  \textstyle )$ is a complex white Gaussian noise with independent real and imaginary parts.\par 
 \par We also notice that in order to ensure the identifiability of the model \eqref{eq:fft} we are working in the parameter space: $\mathbf{A}=\left\{ (\boldsymbol{\alpha^{*}, \theta^{*}, v^{*}}) \in [ -\pi, \pi [^{3\times n}: \alpha_{1}=1, \theta_{1}=0, v_{1}=0  \right\}$. 
 
To summarize, in this section we  estimate the transformation parameters $(\boldsymbol{\alpha^{*}, \theta^{*},  v^{*}})$ without prior knowledge of the function $f$. The estimation depends on the unknown functional parameter $\left(c_{l} \left(f \right) \right)_{\left|l\right| \leq (J-1)/2}$, the Fourier coefficients of the unknown function $f$. So that, we consider a semi-parametrical method based on an $M$-estimation procedure. $M$-estimation theory enables to build an estimator defined as a minimiser of a well-tailored empirical criterion that is given in the following subsection.

\subsection{Parameters estimation procedure} 
The goal is to estimate the vector of parameters  $(\boldsymbol{ \alpha^{*}, \theta^{*},v^{*}})$ that depends on the Fourier coefficients of the unknown function $f$.  We consider a semi-parametric method based on an M-estimation procedure \citep{Fgamb2007}.
\par To construct an M-estimator,  we define the rephased (untranslated and rescaled) coefficients for any vector $(\boldsymbol{\theta, \alpha, v}) \in \mathbf{A}$:
\begin{displaymath}
\tilde{c}_{kl}(\alpha, \theta, v) = \left\{ \begin{array}{ll} \frac{1}{\textstyle \alpha_{k}} e^{il\theta_{k}} d_{kl}, \ \  {1 \leq k \leq n , \ \ 0 <\left|l\right| \leq (J-1)/2 }
\\ \frac{1}{\textstyle \alpha_{k}}\left( d_{kl} - v_{k} \right), \ \  {1 \leq k \leq n ,\ \ l=0 }
 \end{array} \right.
\end{displaymath}

and the mean of these coefficients:
\begin{displaymath}
\hat{c}_{l}(\alpha, \theta, v)=\frac{1}{n}\sum_{k=1}^{n}\tilde{c}_{kl}(\alpha, \theta, v),\ \  {\left|l\right| \leq (J-1)/2}
\end{displaymath}

Therefore, for $(\boldsymbol{\alpha^{*}, \theta^{*} , v^{*}})$ we obtain:
\begin{gather*}
\tilde{c}_{kl}(\boldsymbol{\alpha^{*}, \theta^{*}, v^{*}}) =c_{l}(f)+\frac{1}{\textstyle\alpha^{*}_{k}}\ \ e^{il \theta^{*}_{k}}  w_{kl}\ \ {1 \leq k \leq n} 
\\
\hat{c}_{l}(\boldsymbol{\alpha^{*}, \theta^{*}, v^{*}})=c_{l}(f)+\frac{1}{n} \sum_{k=1}^{n}\frac{e^{il\theta^{*}_{k}} \cdot w_{kl}}{\alpha^{*}_{k}} 
\end{gather*}.

Hence, $\left|\tilde{c}_{kl}(\alpha, \theta, v)-\hat{c}_{l}(\alpha, \theta, v)\right|^{2}$ should be small when $(\boldsymbol{\alpha, \theta, v})$ is closed to $(\boldsymbol{\alpha^{*}, \theta^{*}, v^{*}})$. 
\par Now, consider a bounded positive measure $\mu$ on $[0,T]$ and set
\begin{equation}
\label{eq:delta}
\delta_{l}:=\int^{T}_{0}\exp\left(\frac{2i\pi l}{T}\omega\right)d\mu(\omega),\  \ (l \in \mathbb{Z})
\end{equation}
Obviously, the sequence $(\delta_{l})$ is bounded. Without loss of generality we will assume that $\delta_{0}=0\ \ \text{and}\ \ \left|\delta_{l}\right|>0, l\neq0$. Assume further that $\sum_{l}\left|\delta_{l}\right|^{2}\left|c_{l}(f)\right|^{2}<+\infty$. So that $f\ast \mu$ is a well defined square integrable function: 
\begin{center}
$f\ast \mu (x) = \int f(x-y) d \mu(y)$
\end{center}

We construct the following empirical contrast function (\ref{eq:mn}):
\begin{equation}
\label{eq:mn}
M_n(\alpha, \theta, v)=\frac{1}{n} \cdot \sum_{k=1}^{n} \sum_{l=-\frac{J-1}{2}}^{\frac{J-1}{2}}\left|\delta_{l}\right|^{2} \left|\tilde{c}_{kl}(\alpha, \theta, v)-\hat{c}_{l}(\alpha, \theta, v)\right|^{2}
\end{equation}

The random function $M_{n}$ is non negative. Furthermore, its minimum value is reached closely to the true parameters $(\boldsymbol{ \alpha^{*}, \theta^{*},v^{*}})$. We define the estimator by:
\begin{equation*}
(\widehat{\alpha},\widehat{\theta}, \widehat{v} )_{n}=\arg \min_{{\alpha}, {\theta}, {v} \in \mathbf{A}}M_n(\alpha, \theta, v)
\end{equation*}

 The proof of convergence  $(\widehat{\alpha},\widehat{\theta}, \widehat{v} )_{n}\overset{\mathbf{P}}{\underset{n\to+\infty}{\longrightarrow}}(\boldsymbol{\alpha^{*}, \theta^{*}, v^{*}})$ and asymptotic normality of the estimators can be found in \cite{Fgamb2007, vimond2007}. The weights $\delta_{l}$ in (\ref{eq:delta}) are chosen to guarantee the convergence of the contrast function to a deterministic function $M_n$ and to provide the asymptotic normality of the estimators.   Moreover, the convergence can be speeded up by proper selection of weights. The analysis of covergence at different weights is presented in  \citep{Fgamb2007}. In this work we used the weight $\delta_{l}=1/{\left|l\right|}^{\beta}$ with $\beta=1.5$.
\par
The computation of the estimators is very fast since only a Fast Fourier algorithm and a minimization algorithm of a quadratic functional are needed. The procedure is summarized in Algorithm \eqref{Mest}
\begin{algorithm}
\begin{spacing}{1.2}
\caption{\label{Mest}Parameters estimation procedure}
\begin{algorithmic}
\REQUIRE {Input set of curves from experimental design $\mathbf{Y}^{n}=\{ Y_{ij}, i=1,..,n;\ \ j=1,..,J\}$} 
\ENSURE{Transformation parameters estimation $(\boldsymbol{\alpha^{*}, \theta^{*}, v^{*}})$}
\STATE{Define the identifiability condition: $\mathbf{A}=\left\{ (\boldsymbol{ \alpha^{*}, \theta^{*},v^{*}}) \in [ -\pi, \pi [^{3\times n}: \alpha_{1}=1, \theta_{1}=0, v_{1}=0  \right\}$}
\par
\STATE{Compute the matrix of discrete Fourier coefficients $D=\{d_{kl},\ \ k=1,..,n; \ \ \left|l\right|\leq (J-1)/2\}$ }
\STATE{Compute the matrix of rephased Fourier coefficient $\tilde{C}=\{ \tilde{c}_{kl},\ \ k=1,..,n; \ \left|l\right|\leq (J-1)/2\}$}
\STATE{Compute the vector of mean of rephased coefficients $\widehat{C}=\{ \widehat{c}_{l},\ \ \left|l\right|\leq (J-1)/2\}$}
\STATE{Choose the weight sequence $\delta_{l}$}
\STATE{Define $M_n(\alpha, \theta, v)=\frac{1}{n} \cdot \sum_{k=1}^{n} \sum_{l=-\frac{J-1}{2}}^{\frac{J-1}{2}}\left|\delta_{l}\right|^{2} \left|\tilde{c}_{kl}(\alpha, \theta, v)-\hat{c}_{l}(\alpha, \theta, v)\right|^{2}$ }
\STATE{Compute $(\boldsymbol{\widehat{\alpha},\widehat{\theta},\widehat{v}})=\arg \min_{{\alpha}, {\theta}, {v} \in \mathbf{A}}M_n(\alpha, \theta, v) \in \mathbb{R}^{3\times (n-1)}$ }
\RETURN {$(\boldsymbol{\widehat{\alpha},\widehat{\theta},\widehat{v}})\in \mathbb{R}^{3\times (n-1)}$}
\end{algorithmic}
\end{spacing}
\end{algorithm}

\subsection{Analytical Function Example}
In this section the shape invariant model and the efficient parameters estimation are presented on an analytical function. The minimization algorithm used in the estimation procedure is a Newton-type algorithm.
\par Let us consider the following function:
\begin{equation*}
f(x)=20\cdot (1-x/(2\pi))\cdot x/(2\pi)
\end{equation*} 
Simulated data are generated as follows:
\begin{equation*}
Y_{ij} = \alpha^{*}_{i}\cdot f(t_{j}-\theta^{*}_{i})+ v^{*}_{i}+ \varepsilon_{ij}, \ \ 1\leq i \leq n , 1\leq j\leq J 
\end{equation*}
with the following choice of parameters: $J=5, \ \ N=101, \ \ t_{j}=\frac{j-1}{J} 2 \pi, \ \ 1 \leq j \leq J$ are equally spaced points on $[0, 2\pi[$. Transformation parameters $(\boldsymbol{\theta^{*},\alpha^{*}, v^{*}})$   are uniformly distributed on $]0, 1]$, where $\theta^{*}_{1}=0,\ \ v^{*}_{1}=0,\ \ \alpha^{*}_{1}=1$; the noise $\varepsilon_{ji}, j=1,..,J, i=1,..,N$ are simulated with a Gaussian law with mean $0$ and variance $0.5$.
\par
\begin{figure}[h]
	\centering
\subfigure[Original noisy data]{
	\includegraphics[height=0.4\textwidth]{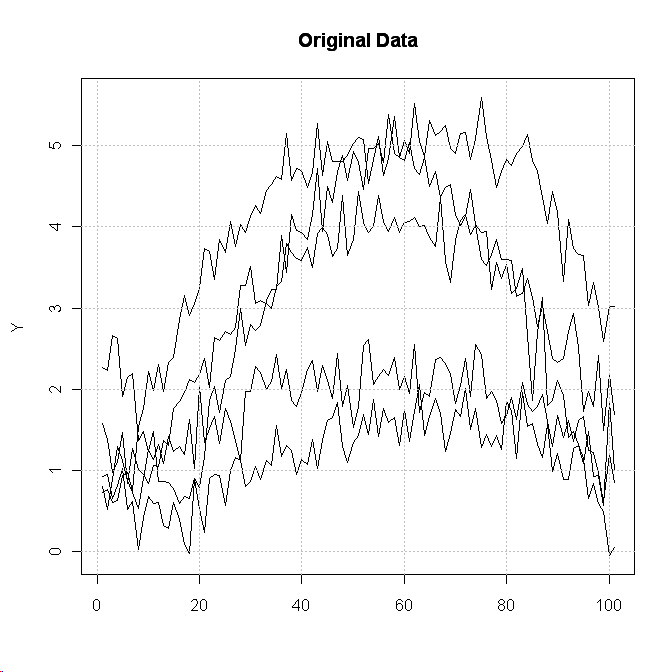}
	\label{fig:original}}
\subfigure[Calculated parameters]{
		\includegraphics[height=0.4\textwidth]{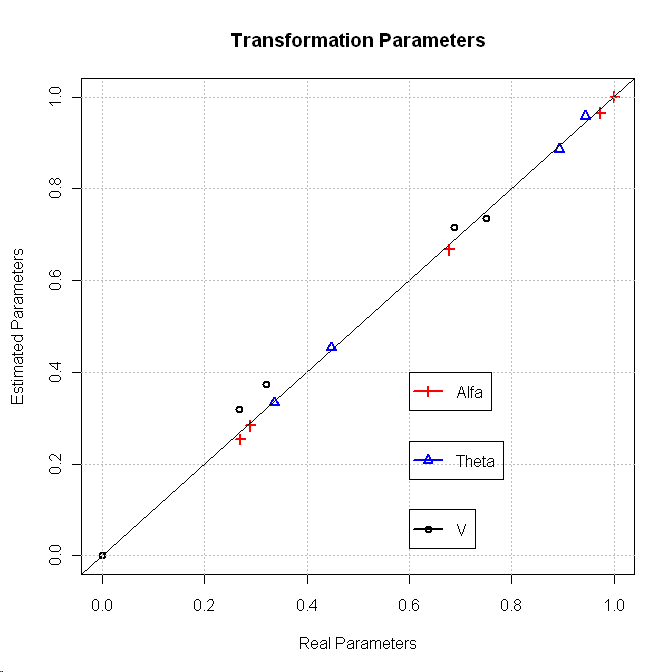}
		\label{fig:params}} 
\subfigure[Inverse Transformation]{
		\includegraphics[height=0.4\textwidth]{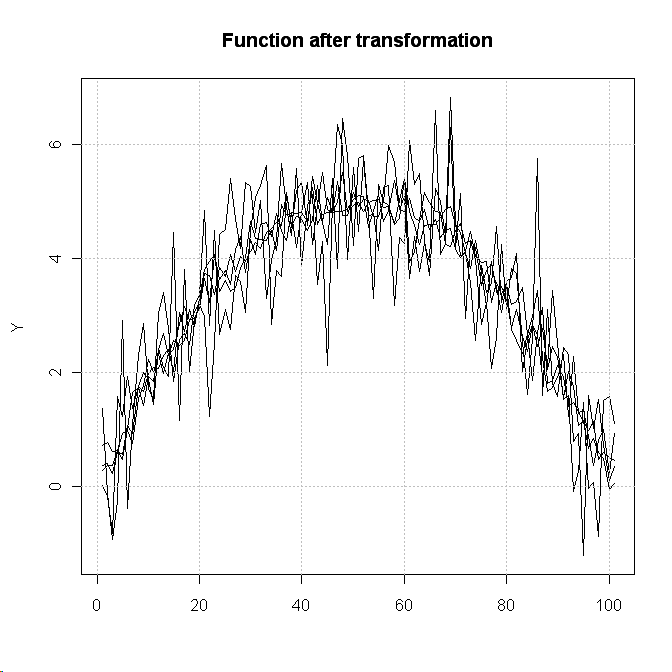}
		\label{fig:trans}}
\subfigure[Comparison]{
		\includegraphics[height=0.4\textwidth]{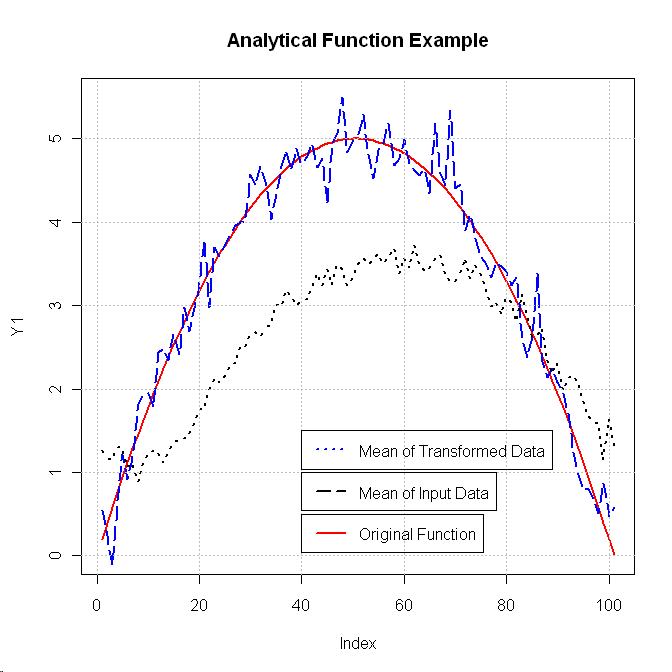}
 		\label{fig:comp}}
	\caption{	\label{fig:funcex}Analytical Example}
\end{figure}

Results are displayed in Figure \ref{fig:funcex}. The function $f$ is plotted by a solid red line in Figure \ref{fig:comp}. Figure \ref{fig:original} shows the original simulated noisy data ${Y_{i,j}}$. The cross-sectional mean function of these data is presented in Figure \ref{fig:comp} by black dotted line. Figure \ref{fig:params} plots estimated transformation parameter versus the originally simulated parameters. As it can be seen, the estimations are very close to the original parameters. The inverse transformation using the estimated parameters is displayed in Figure \ref{fig:trans} and the mean function of restored curves is displayed in Figure \ref{fig:comp} by blue dashed line. Figure \ref{fig:comp} compares the cross-sectional mean of inversely transformed data and the cross-sectional mean of originally simulated data. Despite the noise, it is  noticeable that the data after inverse transformation are much more closer to the original function $f$ than the original cross-sectional mean function. 

This analytical example shows that the method is effective in estimating the transformation parameters of the shape invariant model. In the next section, we will explain how this model can be applied in reservoir engineering forecast problems.

\section{Functional data approximation}
To apply the shape invariant model approach to approximate functional data from a dynamic simulator, firstly we have to modify the parameters estimation procedure for large number of curves. When we are working with uncertainty modeling, we always start from an experimental design $\mathbf{X}^{n}$ and a set of observation $\mathbf{Y}^{n}$. As we 
have mentioned, the number of design points depends on the number of inputs and on the complexity of the response. So that, some optimization problems could arise when we compute the contrast function with a large number of curves. It can be time consuming and the results can be inaccurate. Therefore, we propose the following modification to the Algorithm \eqref{Mest}: the original observation data is split into  blocks and the optimization procedure is then performed on each of the blocks. Following the identifiability condition, we are working on a compact set $\mathbf{A}=\left\{ (\boldsymbol{\alpha^{*}, \theta^{*}, v^{*}}) \in [ -\pi, \pi [^{3\times n}: \boldsymbol{\alpha}_{1}=1, \boldsymbol\theta_{1}=0, \boldsymbol{v}_{1}=0  \right\}$. This condition should be satisfied on every block optimization by adding the first reference curve to the block.

\begin{algorithm}[H]
\begin{spacing}{1.2}
\caption{Parameters estimation procedure for large $n$}
\label{MNest}
\begin{algorithmic}
\REQUIRE{Input set of curves from experimental design $\mathbf{Y}^{n}=\{ Y_{ij}, i=1,..,n;\ \ j=1,..,J\}$} 
\ENSURE{Transformation parameters estimation $(\boldsymbol{\alpha^{*}, \theta^{*}, v^{*}})$}
\STATE{Split the observation data into $N_{b}$ blocks of $(K+1)$ curves}
\FOR {$m=1,..,N_{b}$} 
\STATE{Define block curves $\mathbf{Y}^{K+1}=\{Y_{1},Y_{(m-1)(K+1)...,Y_{mK}}\}=\{Y_{ij},\ \ i=1,..,K+1, j=1,..,J\}$}
\STATE{Perform Algorithm \eqref{Mest}}
\STATE{Compute  $(\boldsymbol{\widehat{\alpha},\widehat{\theta},\widehat{v}})=\arg \underset{\alpha, \theta, v \in \mathbf{A}}{\min}M_n(\alpha, \theta, v)$, where $(\boldsymbol{\widehat{\alpha},\widehat{\theta},\widehat{v}})\in \mathbb{R}^{3\times K}$ }
\ENDFOR 
\RETURN{$(\boldsymbol{\widehat{\alpha},\widehat{\theta},\widehat{v}})\in \mathbb{R}^{3\times (n-1)}$}
\end{algorithmic}
\end{spacing}
\end{algorithm}

With this procedure, we do not have limitations on experimental design of any size. As soon as we have estimated the parameters for every curve from observation data set, we can formulate the prediction algorithm.  
Instead of reproducing the simulator output for a prediction point  $\bar{x}_{0}$ at every time step, we model the whole output curve with the transformation parameter. This curve will provide the approximation of the output for the selected input configuration $\bar{x}_{0}$  at each of considered time steps $\{t_{j},\ \ j=1,..,J\}$. 
The transformation parameters for the input $\bar{x}_{0}$ are evaluated with the Gaussian process response surface modeling. The model is based on the experimental design and the set of evaluated transformation parameters calculated for the observation data curves. The prediction framework for an  arbitrary  input configuration  $\bar{x}_{0}$ is presented by the following Algorithm \eqref{forecastalg}. 
\begin{algorithm}
\begin{spacing}{1.2}
\caption{Prediction algorithm for dynamic simulator}
\label{forecastalg}
\begin{algorithmic}
\REQUIRE{Dynamic simulator $Y=F(\bar{x},t)$ with $t\in \{t_{j},\ \ j=1,..,J\}$ and prediction point $\bar{x}_{0}$ }
\ENSURE{Prediction $Y^{0}=F(\bar{x}_{0},t_{j})$ for all $j=1,..,J$}
\STATE{Generate an experimental design $\mathbf{X}^n=(\bar{x}_1,..,\bar{x}_n)$ to span the space of interest }
\STATE{Evaluate $\mathbf{Y}^{n}=\F(\mathbf{X}^{n}, t_{j})$ at every time step $t_{j},\ \ 1\leq j\leq J$}
\STATE{Generate a set of discrete curves $\{Y_{i,j}\},\ \ i=1,..,n; j=1,..,J$}
\STATE{Estimate the  $(\boldsymbol{\alpha, \theta, v})\in (\mathbb{R}^{n})^{3}$ with Algorithm \ref{MNest}}
\STATE{Construct new experimental design for the function of parameters: $(\mathbf{X}^n, \boldsymbol\theta(\mathbf{X}^n)), \ \ (\mathbf{X}^n, \boldsymbol\alpha(\mathbf{X}^n))$ and $(\mathbf{X}^n, \boldsymbol v (\mathbf{X}^n))$}
\STATE{Estimation of hyperparameters for GP models of transformation parameters}
\STATE{ $\boldsymbol\alpha(\bar{x}_{0}),\ \ \boldsymbol\theta(\bar{x}_{0})$ and $\boldsymbol v(\bar{x}_{0})$  are approximated with corresponding GP models }
\STATE{Reproduce: $F(\bar{x}_{0},t_{j})= \boldsymbol{\alpha}(\bar{x}_{0})  f(t_{j}-\boldsymbol{\theta}(\bar{x}_{0}))+\boldsymbol{v}(\bar{x}_{0})$ for all  $\{t_{j},\ \ j=1,..,J\}$} 
\RETURN{Discrete time series $Y^{0}=F(\bar{x}_{0},t)$ with  $t\in \{t_{j},\ \ j=1,..,J\}$ } 
\end{algorithmic}
\end{spacing}
\end{algorithm}
\par
Summing up, with this proposed algorithm the problem of response surface modeling for dynamic simulators is reduced from single step GP modeling for each of $J$ time steps to an optimization problem and a GP modeling for the transformation parameters. However, it is worth to mention that before performing the algorithm it is important to analyze the curves behaviour for the observation data set. Studying the curves characterization, probably we may fix for example vertical shifts $\boldsymbol{v}$ or horizontal  shifts $\boldsymbol{\theta}$ at zero. Also, if the curves have significantly different behaviour at different time intervals we can split the observation data in time as well in order to achieve higher prediction accuracy.
\par
The next section presents an application with a dynamic reservoir simulator case.
  
\section{$\CO_{2}$ storage reservoir case}
\begin{figure}[H]
\centering
\includegraphics[width=0.55\textwidth]{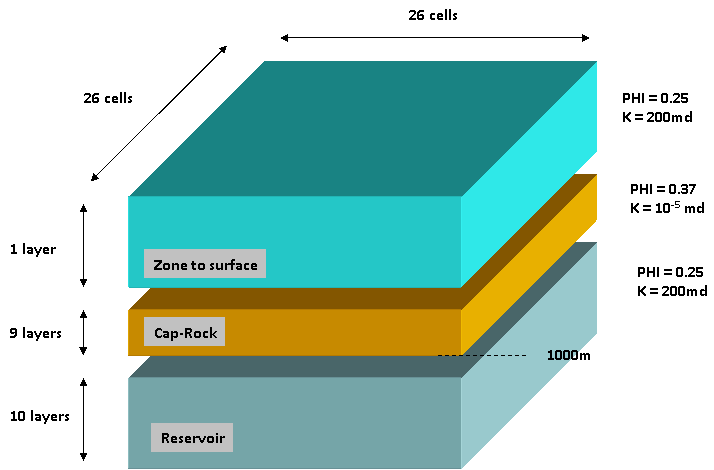}
\caption{\label{co2model}Reservoir Model}
\end{figure}

Carbon Capture and Storage technology stands for the collection of $\CO_{2}$  from industrial sources and its injection underground. Carbon dioxide is stored in a deep geological formation that is sealed on a top by a low permeability cap rock formation. Subsurface storage of $\CO_{2}$ is always associated with an excess pressure in the reservoir and one of the primary environmental risks is a pressure-driven leakage of  $\CO_{2}$ from the storage formation. In order to assess the risk of  $\CO_{2}$ leakage through the cap rock we consider a synthetic reservoir model. The model is made up of three zones \eqref{co2model}:
\begin{itemize}
	\item a reservoir made of 10 layers
	\item a cap-rock made up of 9 layers
	\item a zone-to-surface composed of 1 layer
\end{itemize}

The XY size of the grid is set at 10 km total length. Each layer is 5m thick, including the cell above the cap-rock. The total number of cells is 13520 (26x26x20 model grid). The structure of the reservoir is reduced to its simplest expression. 	The zone above the cap-rock (up to the surface) is currently set to 1 layer. The salinity of the water is 35gm/l. The temperature of the reservoir is set to 60C and the initial pressure is hydrostatic. The injection bottom rate is set to 10E+06 tons/year. The fracture pressure is estimated by geomechanical experts to 158 bars. Exceeding this value of  reservoir pressure can lead to a leakage. The simulation period is 55 years that include an injection period of 15 years followed by 40 years of storage. In this study we analyze the possibility of leakage through a cap rock. Therefore, we consider pressure in the storage reservoir  as an objective function to be approximated.
\par
\begin{figure}[H]
\centering
\subfigure[Observation data]{
	\includegraphics[width=0.45\textwidth]{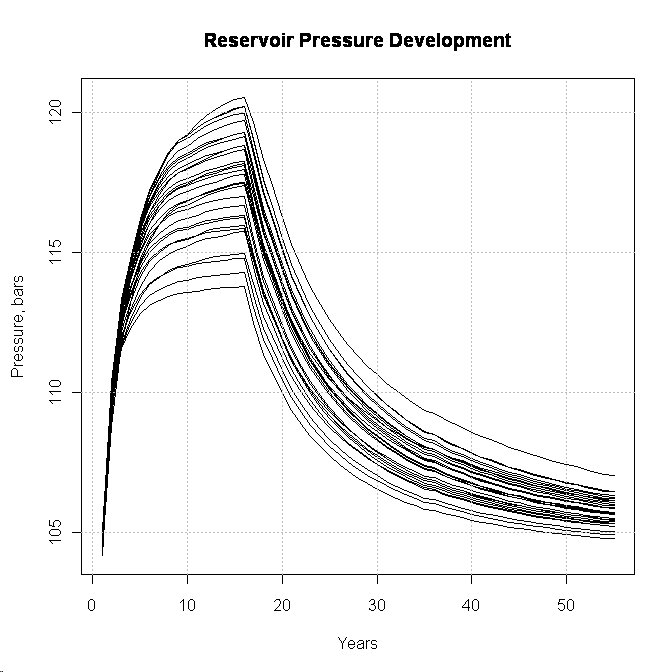}
	\label{fig:co2press}}
  \subfigure[Restored data]{
	\includegraphics[width=0.45\textwidth]{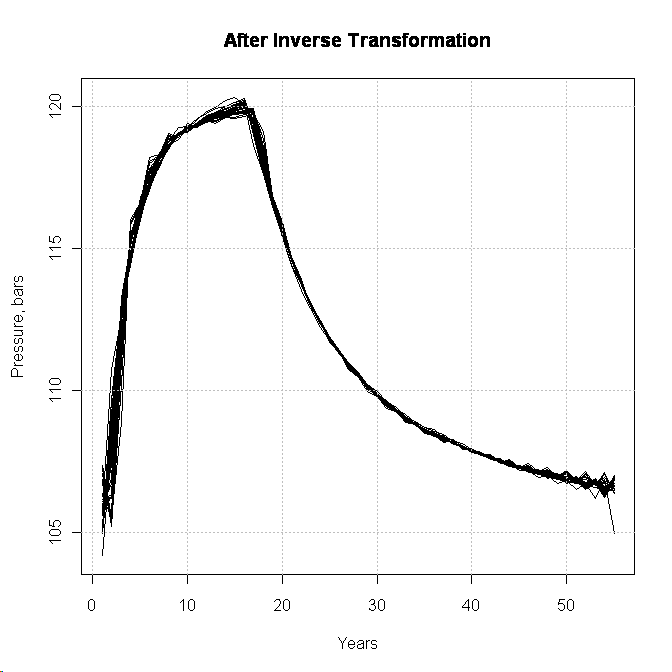}
	\label{fig:CO2pressinv}}
\caption{\label{rescaserestored}Original observation data and data after inverse transformation}
\end{figure}

The uncertain parameters  selected for this study characterize the reservoir and the fluid properties. It implies different $\CO_{2}$ flowing possibilities between the reservoir layers. The distribution law for the parameters is uniform. Table \eqref{tab:UP} represents the parameters description with their range of minimum and maximum values. 
\begin{table}[H]
	  \centering
  	\begin{tabular}{c|l|cc}
		\hline
				Name&Description&Min&Max\\
		\hline
	PORO&Reservoir Porosity&0.15&0.35\\
	KSAND&Reservoir Permeability&10&300\\
	KRSAND&Water relative permeability end-point&0.5&1.0\\
\hline
		\end{tabular}
		\caption{\label{tab:UP}Uncertain Parameters}
\end{table}
 We start from the observation data $\mathbf{Y}^{n}$ - $[30;55]$ matrix of simulator outputs. By means of Algorithm \eqref{MNest} we provide the transformations parameters estimations. Figure \eqref{fig:co2press} provides the original set of curves and Figure \eqref{fig:CO2pressinv} represents the same set after inverse transformation. The pattern curve is differentiable after the inverse transformation with the estimated parameters. 
As we are sure here that the model parameters are efficiently estimated, we can proceed with the next step of prediction algorithm \eqref{forecastalg}. The next step is to build Gaussian process response surface models for the transformation parameters basing on the estimations.
\begin{figure}[ht!]
\centering
\subfigure[RMSE]{
\includegraphics[width=0.45\textwidth]{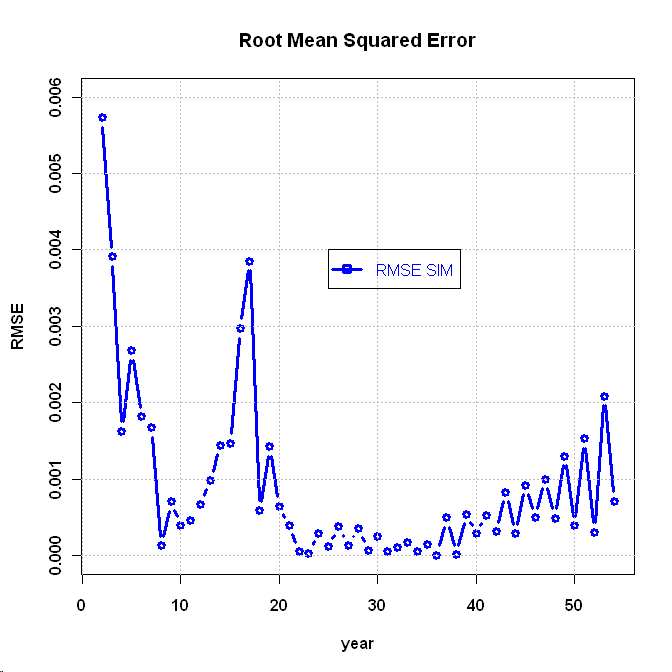}}
\subfigure[Predicitivity Indices]{
\includegraphics[width=0.45\textwidth]{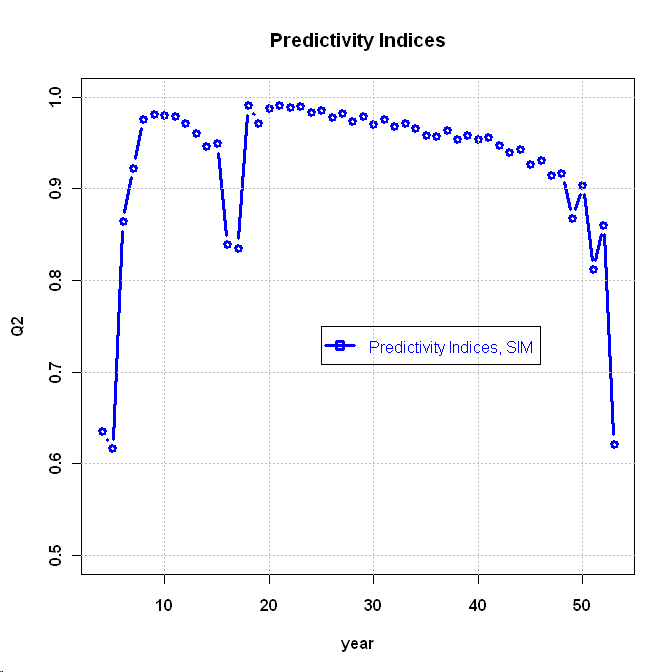}}
\caption{\label{rmse}Predictivity Indices and RMSE}
\end{figure}

\par
\begin{figure}[H]
\centering
\subfigure[SIM method]{
	\includegraphics[width=0.45\textwidth]{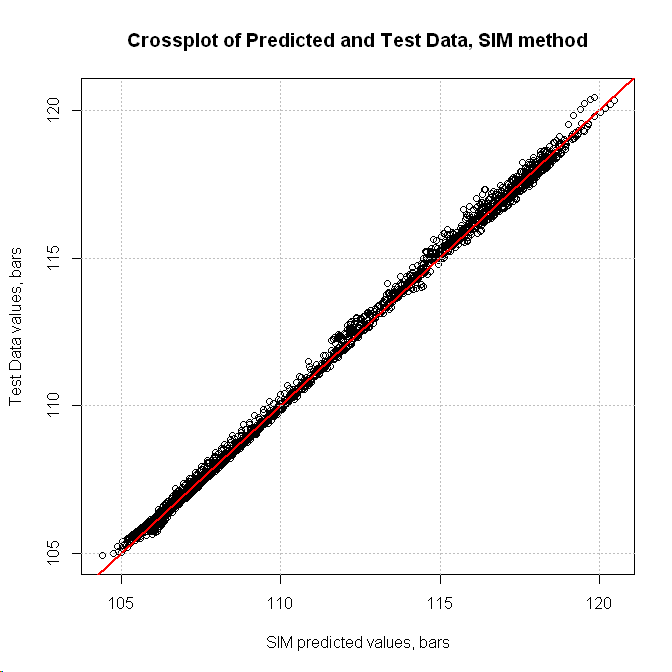}
	\label{fig:sim}}
  \subfigure[Single Step GP]{
	\includegraphics[width=0.45\textwidth]{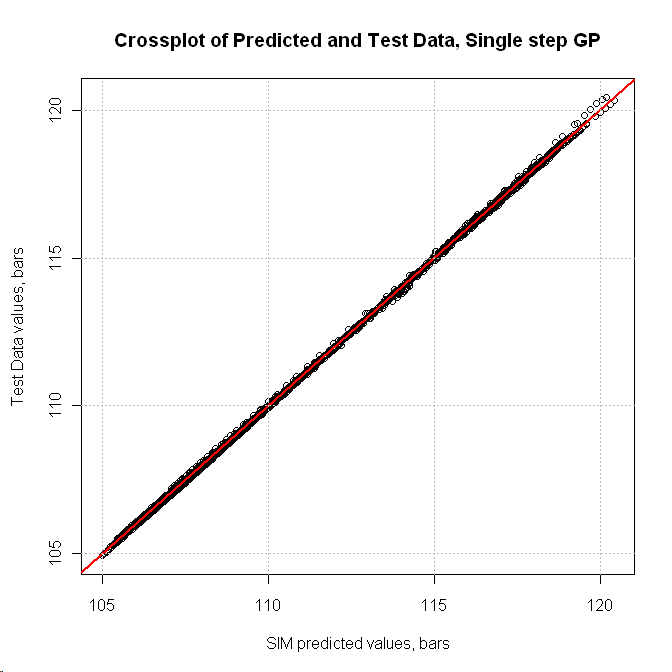}
	\label{fig:gp}}
\caption{Crossplot comparison}
\end{figure}

 In figure \eqref{rmse} we display the model validation criteria: Root Mean Square Error (RMSE) and predictivity indices (Q2), calculated separately for every year. The criteria were computed with the help of additional  confirmation test data. The low predictivity in the first and last years is caused by low variance of data in that period. In general, the method provides reliable level of predictivity. It is also reflected by crossplots of test and predicted data.  Figure \eqref{fig:sim} is based on new proposed approach and Figure \eqref{fig:gp} corresponds to single step GP modelling. Both methods provide a good level of approximation.
Table \eqref{tab:cpu} compares the simulation CPU time for both methods. SIM method provides suffucuently accurate results (although less accurate than the single step GP approach), but with a CPU reduction of a factor five.
\begin{table}[H]
	\centering
		\begin{tabular}{c|c|c|c|c}
	\hline
				& \multicolumn{3}{|c|}{SIM method}&\multirow{2}{*}{Single step GP}\\
  \cline{2-4}
				&Optimization &Parameters modeling&Total&\\
	\hline
CPU time&00:00:35&00:00:15&00:00:50&00:04:28\\
  \hline
		\end{tabular}
		\caption{\label{tab:cpu}CPU time comparison}
\end{table}
It is worth to mention, that in this study we consider a simple model with only 3 uncertain parameters. So that, to estimate the function with GP model at every single step takes approximately 10 seconds. For more complex functions and more input uncertain parameters involved a single step model evaluation can take up to 10-20 minutes. So that, for the same simulation period of 55 years CPU time can increase to 10-20 hours. Whereas SIM approach does not depend on number of time steps and the method always conclude only a single optimization problem and as maximum 3 GP models for the transformation parameters.

\section{Conclusion}
This paper focuses on two general problems. First, we introduce a shape invariant model approach and we provide an efficient algorithm for estimation of transformation parameters of the model with the method specification for large sets of curves. Second, we suggest the application of this approach to model the time series outputs from a dynamic simulator. The proposed method reduces the problem of functional outputs modeling to one optimization problem and three GP response surface models. We have tested the method with a $\CO_{2}$ storage reservoir case. We have also compared the method with the standard single-step approach. Presented numerical results show that the method provides satisfactory and comparable predictivity at lesser CPU time. The method also does not depend on the number of the involved time-steps. It can be very advantageous when we are working with a model involving large number of times steps such as $\CO_{2}$ storage when the reservoir model simulation period can include up to hundreds or thousands time steps. However, if the set of output curves have significantly different behaviour, preliminary curves classification may be required.
  
\section{Acknowledgements}
This work has been partially supported by the French National Research Agency (ANR) through COSINUS program (project COSTA-BRAVA n ANR-09-COSI-015).
I also would like to thanks Nicolas Mourand and Dan BOSSIE CODREANU for the presented $\CO_{2}$ reservoir model.

\bibliographystyle{elsarticle-num} 
\bibliography{references}
%\begin{thebibliography}{00}
%\bibitem[
%% \bibitem must have one of the following forms:
%%   \bibitem[Jones et al.(1990)]{key}...
%%   \bibitem[Jones et al.(1990)Jones, Baker, and Williams]{key}...
%%   \bibitem[Jones et al., 1990]{key}...

%%   \bibitem[\protect\citeauthoryear{Jones et al.}{1990}]{key}...
%%   \bibitem[\protect\astroncite{Jones et al.}{1990}]{key}...
%%   \bibitem[\protect\citename{Jones et al., }1990]{key}...
%%   \harvarditem[Jones et al.]{Jones, Baker, and Williams}{1990}{key}...
%%


% \end{thebibliography}

\end{document}